\documentclass{iopjournal}

\usepackage{amssymb}
\usepackage{amsfonts}
\usepackage{amsmath}
\usepackage{physics}
\usepackage{mathtools}
\usepackage{todonotes}

\usepackage{array}
\usepackage{multirow}
\newcolumntype{C}[1]{>{\centering\arraybackslash}m{#1}}

\usepackage{tcolorbox}

\newcommand{\A}{\mathcal{A}}
\newcommand{\B}{\mathcal{B}}
\newcommand{\XX}{\mathcal{X}}
\newcommand{\YY}{\mathcal{Y}}
\newcommand{\ZZ}{\mathcal{Z}}
\newcommand{\Id}{{\rm 1\hspace{-0.9mm}l}}

\newtheorem{example}{Example}

\begin{document}

\articletype{Paper} 

\title{A Bipartite Quantum Key Distribution Protocol Based on Indefinite Causal Order}

\author{Mateusz Le\'s{}niak$^1$\orcid{0009-0001-0092-2975}, Ryszard Kukulski$^{2}$\orcid{0000-0002-9171-1734}, Paulina Lewandowska$^2$\orcid{0000-0003-1564-7782}, 
Grzegorz Rajchel-Mieldzio{\'c}$^3$\orcid{0000-0003-2652-9482} and Micha\l{} Wro\'nski$^{1, *}$\orcid{0000-0002-8679-9399}}

\affil{$^1$NASK National Research Institute, Department of Cryptology, Warsaw, 01-045, Poland}

\affil{$^2$IT4Innovations, VSB~-~Technical University of Ostrava, 
17.~listopadu 2172/15, 708 33 Ostrava, Czech Republic}

\affil{$^3$BEIT sp.\ z o.o., ul.\ Mogilska 43, 31-545 Krak{\'o}w, Poland}

\affil{$^*$Author to whom any correspondence should be addressed.}

\email{michal.wronski@nask.pl}


\begin{abstract}
We propose a bipartite quantum key distribution (QKD) protocol based on \emph{causal nonseparability}: the presence of a resource---a process matrix---that does not correspond to any definite causal order between two parties. In our protocol, Alice and Bob perform local operations arranged in a ``causal-order guessing game,'' whereby each round yields an 85.35\% probability of matching bits when the communication is undisturbed. This raw matching probability (or equivalently, a $\sim14.65\%$ error rate) is amenable to standard forward error-correction strategies.
We further discuss the practical construction of the QKD protocol using indefinite causal order, where several different scenarios are deeply analyzed.
\end{abstract}

\section{Introduction}\label{sec:Introduction}
Quantum computing has triggered a fundamental shift in cryptography. On the one hand, quantum algorithms such as Shor's factoring algorithm threaten many classical public-key systems~\cite{rivest1978method,hellman1976new,Miller1986}, motivating the design of post-quantum cryptosystems in the NIST PQC project. On the other hand, quantum mechanics also supports revolutionary cryptographic primitives such as quantum key distribution (QKD), which can provide unconditional security against arbitrarily powerful eavesdroppers.

Among the best-known QKD protocols, BB84~\cite{BENNETT20147}, E91~\cite{ekert1991quantum}, and B92~\cite{Bennett_1992} have paved the way toward real-world quantum communication networks. These protocols exploit the fundamental no-cloning principle (for BB84 and B92) or quantum entanglement (for E91) to detect any potential eavesdropping. Subsequent work has refined their security analyses to account for device imperfections and finite sample sizes~\cite{Bennett_1995,Renner_2005}, spurring experimental implementations that push the boundaries of optical and quantum technologies~\cite{Ecker_2019,Sohr_2024}.

In parallel, there is growing interest in \emph{indefinite causal structures}~\cite{oreshkov2012quantum,oreshkov2016causal}, which challenge the usual assumption that quantum operations must occur in a definite sequence---either ``Alice acts, then Bob acts'' or the opposite. Process matrices describing such causal non-separability (CNS) do not admit a well-defined ordering of operations yet still produce well-defined probability distributions. It has been shown that these matrices can lead to advantages in certain \emph{causal games}, going beyond what is possible with ordinary, causally ordered processes.

Here, we leverage this phenomenon to construct a bipartite QKD protocol. We show that if Alice and Bob each apply specific local unitaries and measurements on a shared CNS resource, they can produce strongly correlated outcomes with 85.35\% agreement, which corresponds to a raw quantum bit error rate (QBER) of about 14.65\%. While this QBER is higher than the typical $\sim10\%$ threshold often cited for standard QKD experiments, it still lies below capacity bounds that can be approached with advanced error-correction codes such as low-density parity check (LDPC). Furthermore, we carefully examine finite-blocklength scenarios using the Polyanskiy--Poor--Verd\'u (PPV) bounds~\cite{PPV10} to estimate the required block sizes to achieve extremely low decoding (block) error probabilities, such as $10^{-6}$. With the proposed error-correction method we are able to  tolerate errors that does not exceed the maximal tolerable error rate of about 2\%.

In this work our contributions are as follows:
\begin{itemize}
    \item introducing quantum bit exchange procedure based on causal game with probability of correct bit exchange that is \(p_\text{succ} = \frac{2 + \sqrt{2}}{4} \approx 0.8535\);
    \item providing an example key exchange protocol using the designed procedure and classical error correction codes;
    \item estimating the probability of key compatibility for eavesdropper, using SDP optimization;
    \item providing finite‑blocklength analysis estimating the length of the ideal correction code for the presented protocol;
    \item presenting a working simulation of the protocol using concatenation of correction codes: repetition code and BCH code.
\end{itemize}

In the following, we first outline some background on bipartite QKD protocols and how they typically rely on quantum resources such as superposition or entanglement. We then introduce process matrices and explain the distinction between causally separable and causally non-separable resources. Next, we describe our proposed protocol in a step-by-step manner, including security considerations and potential eavesdropping strategies. Finally, we discuss in greater detail how finite-blocklength effects (and, in particular, the PPV approach) shape the required number of qubits to produce a secure final key of a given length with negligible error.

\section{Overview of bipartite quantum key distribution protocols}\label{sec:Overview of bipartite quantum key distribution protocols}
In the classical domain, no key distribution is perfectly secure, as the message can always be intercepted by an adversary in the middle, usually referred to as Eve. 
To achieve a high level of security, various methods are used, including computations that are believed to be hard to perform but easy to verify their correctness, such as multiplication of large primes. 

In contrast, quantum key distribution  relies on the inherently probabilistic quantum nature of particles. 
Their two main aspects are used in various QKD schemes since their inception in the 80s: quantum superposition~\cite{BENNETT20147} and entanglement~\cite{ekert1991quantum,Bennett_1992}. 
The first of these schemes was the famous BB84, introduced by Bennett and Brassard, which includes the preparation of the states in a random basis chosen by the sender (Alice), then measured in a random basis by the receiver (Bob). 
If the choice of these bases (the computational one or the Hadamard basis) matches, then, the measurement should give perfect agreement with the intended message.
Therefore, by only revealing their choices, and then a small subset of the measurements and the messages, they are able to detect an adversary in the middle, as Eve cannot clone the incoming state due to the non-cloning principle. 

Subsequently, in 1991, Artur Ekert devised a scheme with an entangled pair of particles. 
Due to the nature of quantum entanglement, the measurements' results that can be obtained are perfectly correlated, provided the entanglement is maximal (Bell states). 
However, since the quantum correlations are fundamentally random, it is not possible to predict the results of the measurements. 
In particular, using the entangled pair, it should be possible to detect quantum correlations that extend beyond the classical domain. 
Examples of such correlations are the famous Bell correlations, where the results in certain bases are classically confined below 2. 
However, using the quantum particles it was shown that the results of these measurements can go as high as $2\sqrt{2}$. 
Therefore, if this number is reached after revealing a subset of their measurements, the entanglement between Alice and Bob must have been perfect.
As such, no intermediary adversary might have acted; thus, the message is secure. 

The above protocols rely on principles of quantum mechanics realized with discrete-variable systems, such as photons encoding polarization.
This broad class of protocols, dubbed discrete-variable QKD, encompasses also subsequent setups, such as six-state~\cite{Bruss_1998}, B92~\cite{Bennett_1992}, BBM92~\cite{Bennett_1992_v2}, modified encoding patterns~\cite{Scarani_2004,MubashirKhan_2009}, or protocols based on two-way communication~\cite{Bostrom_2002,Beaudry_2013}. 
For the latter group, the principle of operation is slightly different, as the encoding does not rely on the preparation of the initial state.
Instead, a forward and backward communication is established between Alice and Bob.
The security of two-way communication can be increased as compared to the standard, one-way QKD protocols.
Our protocol can be thought of as an example of two-way QKD, but the key distinguishing feature is that the directionality of the sending of information is not fixed but chosen dynamically.

Notwithstanding, the assumption that the Hilbert space, accessible to a given physical system, is a finite-dimensional one, does not always hold in practice.
This paves the way to another way of thinking about quantum key distribution, via setups of continuous-variable QKD~\cite{Ralph_1999,Cerf_2001,Leverrier_2017}. 
These might be realized in a systems such as the electromagnetic field in a coherent state, which is often a more reasonable assumption than a single-photon approximation used in the discrete-variable QKD. 
Then, the variable that is used for the encoding is the amplitude of the coherent state. 
However, these protocols are harder to compare with their discrete counterpart due to the non-existence of a notion of a qubit. 
Therefore, for the goal of the comparison of the efficacy of the protocols, we shall stick to the discrete-variables ones, as ours also belongs to this group.

A lot of the effort in the domain of QKD has been devoted to studying its security under realistic scenarios -- as no measurement is perfect, we expect some noise~\cite{Bennett_1995,Renner_2005}. 
Therefore, the central question of this research is \emph{how much noise still certifies the security of QKD?}
There have been a couple of major achievements in this field, both from theoretical~\cite{Ecker_2019}, as well as an experimental perspective~\cite{Sohr_2024}.
One of the ideas to increase the noise-resistance includes increasing the dimensions of the subsystems, thereby allowing stronger entanglement to occur~\cite{Cozzolino_2019,Erhard_2020}. 
An alternative line of research is devoted to detaching the security of the protocol from the functioning of the quantum devices.
This is possible by working in the device-independent paradigm, where even the untrusted devices can be secure, provided they pass a test, what is usually called ``self-testing''~\cite{Mayers_1998}. 

For a more in-depth analysis of the state of the art, we refer the reader to excellent reviews~\cite{Pirandola_2020,Yang_2023}.
Unlike the above protocols, which assume a definite ordering of quantum operations (e.g., Alice prepares and sends a state, Bob then measures, etc.), our protocol exploits \emph{causally non-separable} resources~\cite{Rozema_2024}. 
A recent proposal introduces QKD arising from indefinite causal orders~\cite{Spencer-Wood_2023}. 
Notwithstanding, it considers sending a state through such a channel and decoding it via BB84, thereby reducing the necessary resources.
In contrast, our goal is to determine the ordering of the actions, which will lead to a common key. 
We do so in the representation of process matrices that do not factor into a neat ``Alice-then-Bob'' or ``Bob-then-Alice'' structure, introduced in the following section.

\section{Quantum bit exchange based on indefinite causal order}\label{sec:Quantum bit exchange based on indefinite causal order}
In this section we present the background to our protocol. The definition of a process matrix, causal game setting and single bit exchange modeled as a binary symmetric channel is presented.

\subsection{Bipartite process matrices}
This section introduces the formal definition of the process matrix \cite{araujo2015witnessing} that describes the causal correlation between the parties. Let us introduce the following notation. Consider two complex Euclidean spaces and denote them by  $\XX, \YY$. By $\mathrm{L}(\XX , \YY)$ we denote the collection of all linear mappings of the form $A: \XX \rightarrow \YY$. As a shorthand put
$\mathrm{L}(\XX) \coloneqq \mathrm{L}(\XX, \XX).$ 
By $\mathrm{Herm}(\XX)$ we denote 
the set of Hermitian operators. 
 Let us define the operator  $\prescript{}{\XX}{Y}$ as 
 \begin{equation}
 \prescript{}{\XX}{Y}  = \frac{\Id_\XX}{\dim(\XX)} \otimes \tr_\XX Y,
 \end{equation} for every $Y \in \mathrm{L}(\XX \otimes \ZZ)$, where $\ZZ$ is an arbitrary complex Euclidean space. 
 We will also need the following projection operator $L_V$ given by  \begin{equation}\label{proj}
 L_V(W) =  \prescript{}{\A_O}{W} +  \prescript{}{\B_O}{W} - \prescript{}{\A_O\B_O}{W} - \prescript{}{\B_I\B_O}{W}  + \prescript{}{\A_O\B_I\B_O}{W} - \prescript{}{\A_I\A_O}{W} + \prescript{}{\A_O\A_I\B_O}{W},
 \end{equation} 
 where $W \in \mathrm{Herm}(\A_I\otimes \A_O \otimes \B_I \otimes \B_O)$.  	We say that $W \in \mathrm{Herm}(\A_I\otimes \A_O \otimes \B_I \otimes \B_O) $ is a process matrix if it satisfies the following conditions
 	\begin{equation}
 	W \ge 0, \,\,\, W  = L_V(W), \,\,\ \tr(W) = \dim(\A_O) \cdot \dim(\B_O),
 	\end{equation}
 	where the projection operator $L_V$ is defined by Eq.~\eqref{proj}. In the definition above, one party, whose we will call Alice has access to both subsystems $\A_I, \A_O$ that indicate input and output, respectively. The other party, whose we will call Bob has access to subsystems $\B_I, \B_O$ that are defined analogously. 
    
    The set of all process matrices will be denoted by $W^{PROC}$. Let us define a useful notion of \emph{quantum combs} in the following manner\cite{chiribella2009theoretical}.
Mathematically, we say that $W^{A \prec B} \in W^{PROC} $ is  a   process matrix 	representing a quantum comb $A \prec B$ if it   satisfies the following conditions
	\begin{equation}\label{def-comb}\begin{split}
	W^{A \prec B} = W'_{\A_I\A_O\B_I} \otimes \Id_{\B_O}, \\
	\tr_{\B_I}  W'_{\A_I\A_O\B_I} = W''_{\A_I} \otimes \Id_{\A_O}.
	\end{split}
	\end{equation}
In particular, a quantum comb $A \prec B$ determines that Alice's and Bob's operations are performed in causal order. This means that Bob cannot signal to Alice and the choice of
Bob’s instrument cannot influence the statistics
Alice records. 

The convex hull of quantum combs of the form $W^{A \prec B} $ and $W^{B \prec A}$ creates the set of causally separable process matrices denoted by $W^{SEP}$. There are, however, process matrices that do not correspond to a causally separable process and such process matrices are known as  causally non-separable (CNS). The examples of such matrices were provided in~\cite{oreshkov2012quantum, oreshkov2016causal}.  
The set of all  causally non-separable process matrices
will be denoted by $W^{CNS}$. 
As an example of a causally non-separable process matrix consider 
\begin{equation} \label{cns}
W^{\text{CNS}} = \frac{1}{4} \left[\Id_{\A_I\A_O\B_I\B_O} + \frac{1}{\sqrt{2}} \left( \sigma_z^{\A_O} \sigma_z^{\B_I}  \otimes \Id_{\A_I\B_O} +  \sigma_z^{\A_I} \sigma_x^{\B_I} \sigma_z^{\B_O} \otimes \Id_{\A_O} \right) \right], 
\end{equation}
where $\sigma_x^\XX, \sigma_z^\XX$ are Pauli matrices\cite{oreshkov2012quantum}. 

\subsection{Causal game setting}\label{subsec:Causal game setting}
Inspired by\cite{oreshkov2012quantum}, we will consider two players, Alice and Bob, who reside in separate laboratories. Both obtain separately a random bit $a$ for Alice and $b$ for Bob. 
In addition, Bob will have to generate another random bit $b'$ which specifies the next step of the experiment. 
In particular, if $b' = 0$, Bob will have to communicate the bit $b$ to Alice, whereas if $b' = 1$, he will have to guess the bit $a$. We also assume that both parties always attempt to guess the other’s bit: Alice’s guess of Bob’s bit is denoted by $x$, and Bob’s guess of Alice’s bit is denoted by $y$. Then,  their goal is to maximize the probability of a successful guess
\begin{equation}
    p_{\text{succ}} \coloneqq \frac{1}{2} P(x=b|b'=0) + \frac{1}{2} P(y=a|b'=1). 
\end{equation}

Using process matrix from $W^{SEP}$, it can be shown that, assuming fixed causal order, no strategy can allow to exceed the bound $p_{\text{succ}} \le \frac{3}{4}$. 
However, by using a process with a process matrix described by Eq. \eqref{cns}, Alice and Bob can increase the success probability in the following way. 
If Bob wants to read Alice’s bit ($b'=1$), then he
 measures his qubit in $Z$-basis and obtains one of the states $\ket{z_{\pm}}$.
Then, if Alice encodes her bit in the $Z$-basis,  Bob can guess Alice’s bit with probability $P(y=a|b'=1) = \frac{2+\sqrt{2}}{4}$. Similarly, if Bob wants
to send his bit ($b'=0$), he measures in the $X$-basis and Alice measures in the $Z$-basis, to finally obtain
$P(x=b|b'=0) = \frac{2+\sqrt{2}}{4}$. In this way they can achieve 
\begin{equation}
    p_{\text{succ}} = \frac{2+\sqrt{2}}{4} \approx 0.8535,
\end{equation}
larger than the causally separable bound.

\subsection{Quantum bit exchange procedure based on causal game} \label{sec:procedure}
This section addresses the generation of the single bit of the cryptographic key using the causal game.
Alice generates a random bit $a$, while Bob generates two random bits $b$ and $b'$. In the next step, Alice prepares the process described by $W^{CNS}$ given by Eq. \eqref{cns} and sends Bob a qubit associated with system $\B_I$. After this, Alice performs the $Z$-basis measurement and reads bit $x$. Finally, she prepares the state $\ket{a}$ on $\A_O$ subsystem.
 Meanwhile, for the received qubit, Bob checks the value of $b'$. 

If $b' = 1$, he performs the $Z$-basis measurement and reads bit $y$. If $b' = 0$, he performs the $X$-basis measurement to get bit $y$. 
After the measurements, he publicly announces $z = (1-b')y \oplus b$ and $b'$.
Alice encodes $z$ as state $\ket{z}$ and feeds it into $W^{CNS}$ in the subsystem $\B_O$.
Alice determines the value of her key $K^A$ in the following way. She reads the value of $b'$. If $b' = 1$, then $K^A = a$. If $b' = 0$, then $K^A = x$. Meanwhile, for Bob's shared key $K^B$, he defines $K^B = y$ if $b' = 1$ and $K^B = b$ if $b' = 0$.

Let us analyze the probability that the bits of $K^A$ and the bit of $K^B$ match. If $b' = 1$, then according to \cite{oreshkov2012quantum} it holds \begin{equation}
    P(K^B = K^A) = P(y = a) = \frac{2+\sqrt{2}}{4}.
\end{equation} Otherwise, if $b'_i = 0$, then we have \begin{equation}
    P(K^B = K^A) = P(x=b) = \frac{2+\sqrt{2}}{4}. 
\end{equation}
That means that keys $K^A$ and $K^B$ comply with the percentage $\frac{2+\sqrt{2}}{4} \simeq 0.8535$.

\subsection{Security analysis of quantum bit exchange round procedure}\label{subsec:Security analysis of quantum bit exchange round procedure}
In this section, following~\cite{renner2008security}, we consider how imperfections or Eve influence can affect the security and reliability of the proposed QKD protocol. Generally, all kind of unwanted noise, imperfections of quantum operations, quantum state transfer, preparation of the process $W^{CNS}$ and finally the influence of Eve can reduce the probability that Alice and Bob keys are compliant. In order to avoid a significant overhead in the number of bits transmitted to establish the correct key, it is important to determine the value acceptance level \(Q_0\). The value will allow the design of a protocol that is resistant to small disturbances.

In this work we focus on security against collective attacks. Extending the bound to fully coherent attacks is left for future work.

\subsubsection{Threat model}
The security of bit exchange is related to how much information Eve can get about $K^A$ or $K^B$. We denote key that Eve defines based on the collected information as \(K^E\). In our threat model Eve has two goals:
\begin{itemize}
    \item maximizing probability \(P(K^E=K^A)\) or \(P(K^E=K^B)\);
    \item staying undetected, the probability of Alice and Bob correctly establish the key stay at acceptance level Q:
    \begin{equation}
        P(K^A = K^B) \geq Q, Q \in \left[\frac12, \frac{2+\sqrt{2}}{4} \right).
    \end{equation}
\end{itemize}
In her best-case scenario (the worst-case scenario for Alice and Bob), every apparatus works perfectly and she is the only factor of disturbance.

We are considering the following scenario:
\begin{enumerate}
    \item Eve intercepts the qubit sent from Alice to Bob on the subsystem $\B_I$;
    \item Eve applies a channel $C_E$ which returns a state on a composed system $\B_I \otimes \mathcal{E}$;
    \item Eve transmit state of $\B_I$ to Bob and store the state of $\mathcal{E}$ for later use;
    \item After Bob announces $z$ and $b'$, Eve performs a binary quantum measurement $\Omega_{z,b'}$ on $\mathcal{E}$ and receives a label $e \in \{0,1\}$, which is her key $K^E=e$. 
\end{enumerate}

Composite strategy of Eve can be expressed as a collection of matrices $\{E_{e,z,b'}\}_{e,z,b'} \subset \mathrm{L}(\B_I \otimes \mathcal{E})$, satisfying quantum network conditions:
\begin{equation}
    \begin{cases}
        E_{e,z,b'} \ge 0, \\
        \sum_e E_{e,z_1,b'_1} = \sum_e E_{e,z_2,b'_2}, \\
        \tr_2(\sum_e E_{e,z,b'}) = \Id_{\B_I}
    \end{cases}
\end{equation}
where the first and the third condition guarantee valid probability distribution generation through positivity and normalization. The second condition states that the results of Eve measurement will not affect statistics of Alice and Bob. Using this notation we can express:
\begin{equation}
    P(x,y,e|a,b,b') = \text{tr} \left( W^{CNS} \left( A_{a,x} \otimes E_{e,z,b'} \star B_{b,b',y} \right) \right),
\end{equation}
where the operator $\star$ is a link product~\cite{bisio2016quantum} defined as \begin{equation}
A \star  B \coloneqq \tr_{\ZZ} \left[ (\Id_\YY \otimes B^{T_\ZZ}) (A \otimes \Id_\XX) \right],
\end{equation}
where $A \in \mathrm{L}(\ZZ \otimes \YY )$, $B \in \mathrm{L}(\XX \otimes \ZZ)$ and $B^{T_\ZZ} $ denotes the partial transposition of $B$ on the subspace $\ZZ$. Then, we use this to write
\begin{equation}
\begin{split}
P(K^A=K^B) &= \frac18 \sum_{a,b} \left( \sum_{y,e} P(b,y,e|a,b,0)+\sum_{x,e} P(x,a,e|a,b,1) \right),\\
P(K^A=K^E) &= \frac18 \sum_{a,b} \left(  \sum_{x,y}P(x,y,x|a,b,0)+ \sum_{x,y} P(x,y,a|a,b,1) \right),\\
P(K^B=K^E) &= \frac18 \sum_{a,b} \left( \sum_{x,y} P(x,y,b|a,b,0)+ \sum_{x,y} P(x,y,y|a,b,1) \right).
\end{split}
\end{equation}

\subsubsection{Estimating the efficiency of Eve} 
\begin{figure}[ht]
    \centering
    \includegraphics[width=0.7\linewidth]{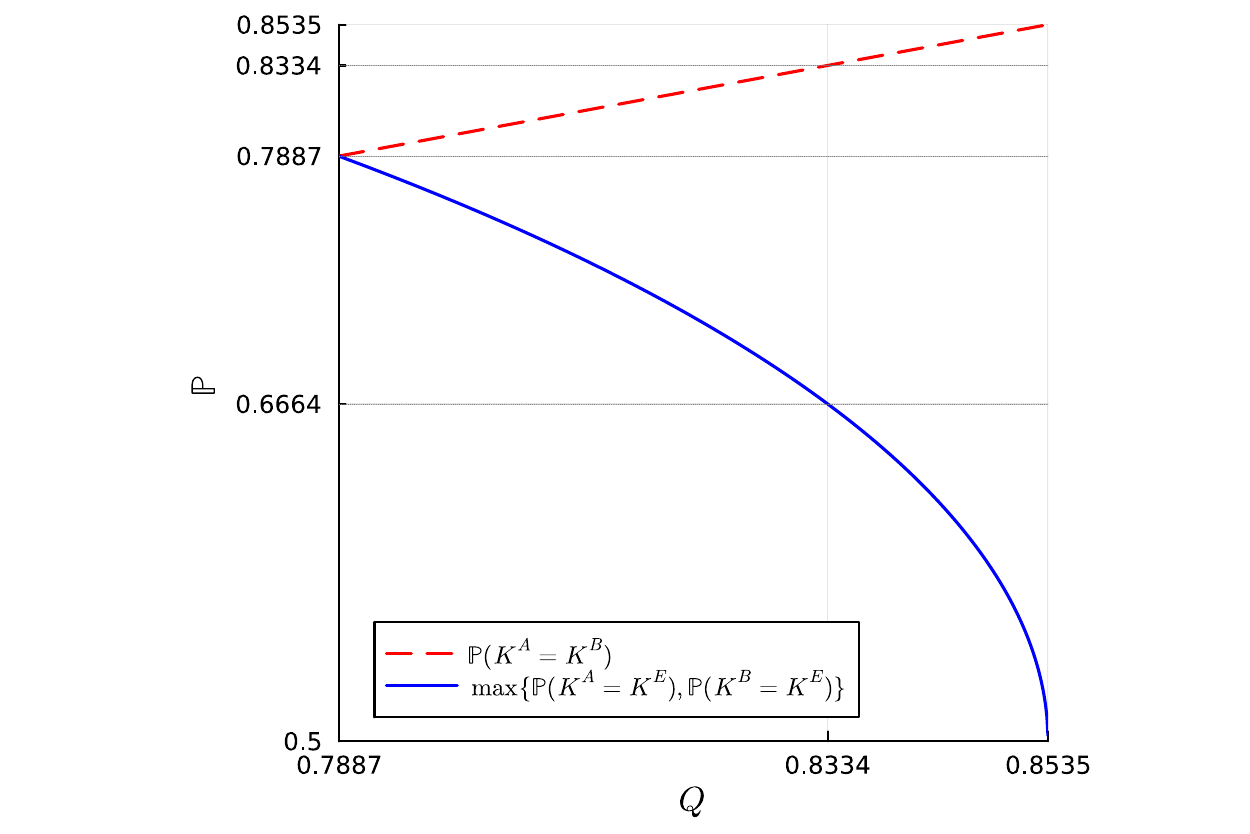}
    \caption{The figure shows the results of SDP optimization. On the $x$ axis we have the acceptance level as a parameter $Q$ and on the $y$ axis we have the probability $\mathbb{P}$ of key compliance for Alice and Bob (dashed red line) and for Eve and Alice (or Bob) (solid blue line). }
    \label{fig:process_matrices}
\end{figure}

To estimate the best probability of compatibility for Eve, we need to solve the following optimization problem:
\begin{equation}
    \max_{X\in\{A,B\}}\max_{E} \{ P(K^E=K^X): P(K^A=K^B) \ge Q\}.
\end{equation}
To determine the solution we used the \texttt{Julia} programming language along with quantum package \texttt{QuantumInformation.jl}\cite{Gawron2018} and SDP optimization via SCS solver~\cite{ocpb:16,scs} with absolute convergence tolerance $10^{-5}$. The code is available on GitHub~\cite{solution}. The figure \ref{fig:process_matrices} shows the results of the computation.

We denoted the probability of Alice and Bob's bits matching with a red dotted line. The solid blue line denotes the greater of the two probabilities of the intercepted bit matching either Alice's bit or Bob's bit. The important value is \(Q \approx 0.7887\), the point where both lines intersect. Starting from this point  Alice and Bob achieve greater key synchronization than Eve. We set the acceptance level for our protocol to $Q_0 = 0.8334$, to find a compromise between accepting the noise and the information about the amount of information that Eve can gain about the secret key. Such a value of acceptance level gives $\max(P(K^E=K^A), P(K^E=K^B)) \le 0.6664$, which means, no more than $2/3$ of the key is known to Eve, whereas Alice and Bob compliance is on the level of $83.34\%$. The key point in the design of the entire protocol is that a valid key can only be obtained for compliance at a level at least equal to the level of acceptability.

Eventually, setting $Q_0$ below the optimal value $\frac{2+\sqrt{2}}{4} \approx 85.36 \%$ means that our protocol is secure and tolerates some amount of errors. Let us model all imperfections as a post-processing bit flip with some error rate $r$. The probability that Alice and Bob keys match is given then as $P(K^A=K^B)(1-r)+P(K^A\neq K^B)r$. Within acceptance level of $Q_0 = 0.8334$, our QKD protocol can tolerate errors with the rate up to $r \le 2 \%$.

\subsection{Modeling as binary symmetric channel}\label{sec:modeling as BSC}
\begin{figure}[ht]
        \centering
        \includegraphics[width=0.5\linewidth]{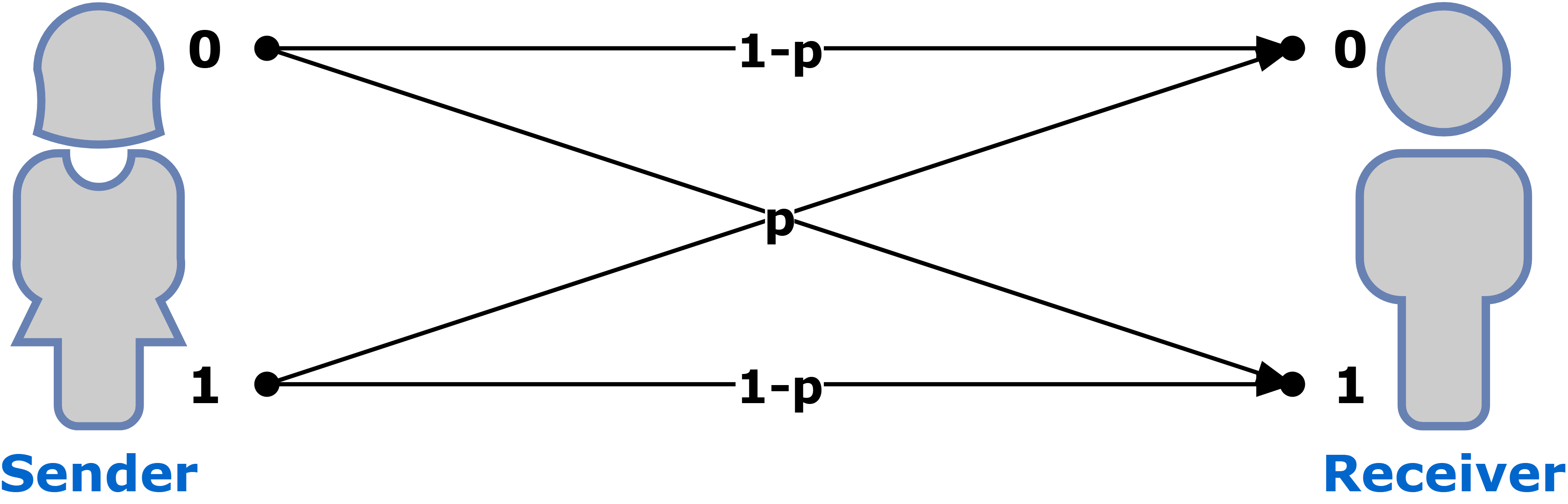}
        \caption{Binary symmetric channel scheme.}
        \label{fig:binary symmetric channel}
    \end{figure}

    Analyzing the procedure described in Section \ref{sec:procedure}, we can see that it works in two different ways:
    \begin{itemize}
        \item if \(b' = 1\), Alice sends her bit \(a\) and Bob receives the same bit with probability \(1-p\);
        \item if \(b' = 0\), Bob sends his bit \(b\) and Alice receives the same bit with probability \(1-p\).
    \end{itemize}
    This brings our procedure to transmitting bit with error between parties. In such situation we can model procedure as two binary symmetric channel whose main idea is presented in Figure \ref{fig:binary symmetric channel}. 
    For the rest of the paper, we denote binary symmetric channel as \(\texttt{BSC}_p(m)\) and define as:
    \begin{equation*}
        \texttt{BSC}_p(m) = \begin{cases}
            m \oplus 1, \text{ with probability } p, \\
            m  \text{ with probability } 1-p.
        \end{cases}
    \end{equation*}
    Depending on \(b'\), the parties send or receive bit through the channel. That means that the direction of our channel depends on value \(b'\) and there is no possibility of exchanging bits in two ways at the same time. 
    This allows us to further simplify our procedure with the use of two-way controlled binary symmetric channel:
    \begin{equation*}
        \texttt{BSC}_p(a, b, b') = \begin{cases}
            \texttt{BSC}_p(a) \text{ if } b' = 1,\\
            \texttt{BSC}_p(b) \text{
            if } b' = 0.
        \end{cases}
    \end{equation*}
    For the rest of the paper, we model quantum bit exchange procedure as described above. Despite the modification we adopted, properties of the classical binary symmetric channel are preserved.

    The channel can be characterized using two parameters: capacity and dispersion. Capacity is defined as follows:
    \begin{equation}
        \label{eq:capacity}
        C = 1 - h(p),
    \end{equation}
    where \(p\) is probability of error and \(h(p) = -p \log p - (1-p) \log (1 - p)\) is binary entropy function. 
    As dispersion of channel we denote:
    \begin{equation}
        \label{eq:dispersion}
        V =p \cdot (1-p) \cdot \Big[\log\frac{1-p}{p}\Big]^2.
    \end{equation}
    Each channel is limited by the maximum cardinality of a codebook of blocklength \(n\) which can be decoded with block error probability greater than \(\epsilon\), denoted as \(M^*(n, \epsilon)\) \cite{Polyanskiy2009}. We can estimate reliably decodable payload size as:
    \begin{equation}
        \label{eq:payload}
        \log M^*(n, \epsilon) = nC - \sqrt{nV} \cdot Q^{-1}(\epsilon) + \frac{1}{2} \log n.
    \end{equation}

    Application of the proposed model simplifies dealing with the error rate in the established key. It reduces our problem to the common problem of error correction \cite{errorcorrectioncodes}, see Section \ref{sec:Finite-blocklength sizing of the reliability code} for a brief discussion of selecting error correction code. 

\section{Quantum key exchange protocol}\label{sec:Quantum key exchange protocol}
In this section we provide description of proposed protocol. The protocol proceeds in three distinct phases: setup phase, quantum bit exchange and finalizing protocol. 

\subsection{Notation}
    To describe our protocol we introduce the following designations:
    \begin{itemize}
        \item \(\texttt{ECC}_{N,K}\) -- error correction code with \(K\)-bit input word and \(N\)-bit output codeword;
        \item \(\texttt{ECC}_{N,K}(\cdot)\) -- encoding of the specified input 
        \item \(\texttt{ECC}^{-1}_{N,K}(\cdot)\) -- decoding of the specified input
        \item \(n\) -- length in bits of private component of established key, must be multiples of \(K\);
        \item \(l\) -- length in bits of encoded private component of established key, \(l =\frac{n}{K} \cdot N\);
        \item \(seed_A\), \(seed_B\) are \(n\)-bit random sequences.
    \end{itemize}

\subsection{Protocol Description}\label{sec:Protocol Description}
    In Table \ref{tab:scheme} our protocol is presented. Setup phase and finalizing exchange are done only once. Quantum rounds are repeated for each established bit. To determine the entire key it takes \(2 \cdot l + \Delta\) rounds. The estimation of the value of \(\Delta\) is shown in Section \ref{subsec:Transmission length and theoretical efficiency}.
    Below we provide a detailed description of each phase. 
    \begin{table}[ht!]
    \centering
    \caption{Schematic representation of the proposed protocol}
    \label{tab:scheme}
    \renewcommand{\arraystretch}{2}%
    \begin{tabular}{|C{0.1\textwidth}|C{0.35\textwidth}C{0.1\textwidth}C{0.35\textwidth}|} \hline
        & \textbf{Alice} & & \textbf{Bob} \\ \hline
        \multirow{ 2}{*}{\textbf{Setup phase}} & \(\texttt{seed}_A \overset{\$}{\gets} \{0, 1\}^n\) & & \( \texttt{seed}_B \overset{\$}{\gets} \{0, 1\}^n\) \\ 
        & \(\texttt{seed}_A^{ECC} \gets \texttt{ECC}_{N,K}(\texttt{seed}_A)\) & & \(\texttt{seed}_B^{ECC} \gets \texttt{ECC}_{N,K}(\texttt{seed}_B) \) \\ \hline \hline
        & \multicolumn{3}{c|}{\textbf{Round} \(i\)} \\ \hline
        \multirow{ 2}{*}{\textbf{1.}} & \multirow{ 2}{*}{\(a \overset{\$}{\gets} \{0, 1\}\)} & & \(b \overset{\$}{\gets} \{0, 1\}\) \\
        & & & \(b' \overset{\$}{\gets} \{0, 1\}\) \\

        \multirow{2}{*}{\textbf{2.}} & \multicolumn{3}{c|}{ \( \texttt{BSC}_p(a, b, b') \rightarrow y_i\)} \\
 
         & \multicolumn{3}{c|}{\(x_i \leftarrow \texttt{BSC}_p(a, b, b')\)} \\
        
        \textbf{3.} & Store value \(x_i\) depends on \(b'\) & & Store value \(y_i\) depends on \(b'\) \\ 
        
        \textbf{4.} & \multicolumn{3}{c|}{Parties check if they collect enough bits to encode current codeword} \\ \hline

        &\multicolumn{3}{c|}{\textbf{Alice collected enough bits and determines the permutation}}\\ \hline
        
        \textbf{5.} & \multicolumn{3}{c|}{\Large \(
            \underset{\longrightarrow}{\Pi_A}\)} \\ 
        
        \textbf{6.} & &  & Retrieves \texttt{codeword} based on received permutation \(\Pi_A\) and stored values \\ 
        
        \textbf{7.} & \(A_j \gets \texttt{ECC}^{-1}_{N,K}(\texttt{seed}_A^{ECC}[j])\) &  & \(A'_j \gets \texttt{ECC}^{-1}_{N,K}(\texttt{codeword})\) \\ \hline

        & \multicolumn{3}{c|}{\textbf{Bob collected enough bits and determines the permutation}} \\ \hline 

        \textbf{5.} & \multicolumn{3}{c|}{\Large \(
             \underset{\longleftarrow}{\Pi_B}\)} \\ 

        \textbf{6.} & Retrieves \texttt{codeword} based on received permutation \(\Pi_B\) and stored values  &  & \\ 
        
        \textbf{7.} & \(B'_j \gets \texttt{ECC}^{-1}_{N,K}(\texttt{codeword})\) & & \(B_j \gets \texttt{ECC}^{-1}_{N,K}(\texttt{seed}_B^{ECC}[j])\) \\ \hline \hline
        \textbf{Privacy amplification} & \multicolumn{3}{c|}{Parties performs privacy amplification and mix halves of the designated keys} \\ \hline
    \end{tabular}
\end{table}

    \subsubsection{Setup phase}
    In this phase Alice and Bob prepare to establish a key. Parties choose theirs private component of the common key and prepare it for transmission.
    \begin{enumerate}
        \item Alice and Bob randomly choose the private component of the common key: \(seed_A, seed_B\);
        \item Alice and Bob encode their components using a publicly known error correcting code and receive \(seed_A^{ECC}, seed_B^{ECC}\). Encoded sequences contain \(\frac{n}{K}\) codewords and are \(l\)-bit long;
        \item Alice and Bob count 1's and 0's in each codeword. They store a list containing pairs of \((1_j, 0_j)\). The parties do not exchange information on the counted numbers.
    \end{enumerate}

    \subsubsection{Quantum bit exchange round}
    In this phase, Alice and Bob establish the bit according to the procedure described in Section~\ref{sec:procedure}.
    To establish a common key, the parties repeat the described round. The proposed round is a combination of quantum process and classical communication. This approach allows to apply the advantages of each part. We introduce the following designations: \(i\) -- index of current round, \(j\) -- index of currently transmitted codeword. Each round consists of the following steps:
    \begin{enumerate}
        \item Alice and Bob prepare for the quantum bit exchange procedure. Parties choose data bits \(a\) and \(b\) they want to exchange, additionally Bob chooses the direction of transmission \(b'\);
        
        \item Alice and Bob perform the procedure described in Section \ref{sec:Quantum bit exchange based on indefinite causal order}. They obtain the same bit with probability \(1-p\). In the case of eavesdropping, the probability of successful transmission of a bit decreases. Due to the selection of the correction code to a specific error level, this will allow the detection of eavesdropping in further steps;
        
        \item The value of \(b'\) determines the direction of the exchange. We assume that the sender knows the exact value of the exchanged bit and the receiver knows the bit with error probability \(p\). Depending on the direction selected, the parties store the index of the current round \(i\):
        \begin{itemize}
            \item if \(b' = 1\): Alice is sender, Bob is receiver;
            \item if \(b' = 0\): Alice is receiver, Bob is sender.
        \end{itemize}  
        The parties then perform the following steps
        \begin{itemize}
            \item Sender checks the value of the bit, and stores \(i\) in the correct set \(\mathbb{O}_j\) or \(\mathbb{I}_j\). 
            Sender knows this bit exactly; therefore, in each round, the value is stored in proper set;
            \item Receiver stores pair \((i: y_i)\) or \((i: x_i)\). The receiver may receive a flipped bit; however, the number of errors should not exceed the decoding radius of the chosen error correction code;
        \end{itemize}
        
        \item Parties check if they sent the entire codeword being currently transmitted:
        \begin{equation*}
            (|\mathbb{O}_A|, |\mathbb{I}_A|) = (0_j, 1_j),
        \end{equation*}
        \begin{equation*}
            (|\mathbb{O}_B|, |\mathbb{I}_B|) = (0_j, 1_j).
        \end{equation*}
        If for neither side the condition is met then they proceed to the next round. If enough bits are collected, the party ready for transmission informs the other party via open channel and prepares a permutation. For each bit \(c_k\) in \(\texttt{codeword}_j\) sender checks
        \begin{itemize}
            \item if \(c_k = 0\), then sender pulls a random index from \(\mathbb{O}\) and appends it to \(\Pi\);
            \item if \(c_k = 1\),  then sender pulls a random index from \(\mathbb{I}\) and appends it to \(\Pi\).
        \end{itemize}
        Once the entire permutation is designated, the sender broadcasts \(\Pi\), without revealing any values of the bits. After broadcasting the permutation, sets storing sender's indices may be discarded. 

        To better illustrate this step of the protocol we provide a pen-and-paper example below.

        \begin{tcolorbox}[
        boxrule=0pt,
        sharp corners]
        \begin{example}\label{example:sending_permutation}
            Parties choose \([7, 4] \)-BCH Code (Bose–Chaudhuri–Hocquenghem codes), while the probability of error is \(\frac{1}{7}\). 
        
        Parties have already performed a certain number of bit exchanges. Bob wants to transfer codeword:
        \begin{equation*}
            \texttt{codeword}_0 = (0, 0, 1, 1, 0, 1, 0).
        \end{equation*}
        He collected required number of 0's and 1's:
        \begin{equation*}
            \begin{array}{c}
                 \mathbb{O}_B = \{0, 2, 7, 13\}, \\
                 \mathbb{I}_B = \{3, 6, 9, 12\}.
            \end{array}
        \end{equation*}
        Then, bit by bit, he analyzes the codeword. First bit of codeword is \(c_0 = 0\) so he pulls a random index from \(\mathbb{O}_B\).
        In our example, he drew~2. Next bit is \(c_1 = 0\), so he pulls index from \(\mathbb{O}_B\) again, reading 0. Third bit to transmit is \(c_2 = 1\), so Bob draws an index from \(\mathbb{I}_B\) and received 3. 
        He then analogously proceeds for the rest of the bits. Finally, he obtained a permutation:
        \begin{equation*}
            \Pi = (2, 0, 3, 9, 7, 12, 13).
        \end{equation*}
        This permutation is sent to Alice.
        \end{example}
        \end{tcolorbox}
        
        \item After receiving permutation, receiver must recover the codeword from it.
        He has a list of pairs in the following form: 
        \begin{equation*}
            \texttt{dict} = \{(0: x_0), (1: x_1), \ldots, (N + \delta: X_{N + \delta})\},
        \end{equation*}
        where \(\delta\) is a small overhead resulting from the random choice of transmitted bits.
        Each element of the permutation maps to a bit from the list. For each index in permutation, select the appropriate item from the list.
        The designated codeword has the following form:
        \begin{equation*}
            \texttt{codeword} = (\texttt{dict}[\Pi(0)], \texttt{dict}[\Pi(1)], \ldots, \texttt{dict}[\Pi(N)]).
        \end{equation*}
        It is worth noting that not all pairs on the list will be used. After designating the codeword, the list of pairs and permutations may be discarded.
        \begin{tcolorbox}[
        boxrule=0pt,
        sharp corners]
        \begin{example}\label{example:decoding_Alice}
            Continuing with Example~\ref{example:sending_permutation} from the previous step, Alice received the permutation and the list of collected pairs:
            \begin{equation*}
                \texttt{dict} = \{(0: 1), (2: 0), (3: 1), (6: 1), (7: 0), (9: 1), (12: 1), (13: 0)\}
            \end{equation*}
            She prepares a codeword; the first element of the permutation is \(\Pi_0 = 2\), which maps to 0. Next element is \(\Pi_1 = 0\), so it maps to 1.
            Alice repeats this analogously for each element of the permutation and obtains the entire codeword:
            \begin{equation*}
            \texttt{codeword}_0 = (0, 1, 1, 1, 0, 1, 0).
            \end{equation*}
        \end{example}
        \end{tcolorbox}
        
        \item In the last step, parties need to decode codewords. The sender of the permutation knows the exact codeword, so is always able to decode the proper part of the key. The receiver has a codeword with \(p \cdot N\) errors. 
        Provided an error correction code was selected, suitable for the specific error rate, both the sender and the receiver should obtain the same word. After decoding, the receiver may check the proper codeword and count errors.

        \begin{tcolorbox}[
        boxrule=0pt,
        sharp corners]
        \begin{example}
            Using values from previous Examples~\ref{example:sending_permutation} and~\ref{example:decoding_Alice}, Alice obtained a codeword, while Bob already knows his codeword. Therefore, they can perform decoding:
        \begin{equation*}
            \begin{array}{cc}
                Alice: & \texttt{ECC}_{7,4}^{-1}((0, 1, 1, 1, 0, 1, 0)) = (0, 0, 1, 0), \\
                Bob: & \texttt{ECC}_{7,4}^{-1}((0, 0, 1, 1, 0, 1, 0)) = (0, 0, 1, 0).
            \end{array}
        \end{equation*}
        The raw codewords differed in one position.
        Nonetheless, due to the error capacity of the selected error correction code, the parties established a common word. 
        \end{example}
        \end{tcolorbox}
        
        \item Parties check if they collected a sufficient number of words; otherwise, they perform the next round.
    \end{enumerate}

    \subsubsection{The final step}
        In the final step, the parties perform privacy amplification, then mix the received components. Alice and Bob splits obtained sequences into \(\texttt{Alice}_L, \texttt{Alice}_U\) and \(\texttt{Bob}_L, \texttt{Bob}_U\), each has \(\frac{n}{2}\) bits. To obtain a common secret key, parties merge corresponding halves, by concatenating them:
        \begin{eqnarray*}
            K_0 = \texttt{Alice}_L || \texttt{Bob}_L\\
            K_1 = \texttt{Alice}_U || \texttt{Bob}_U
        \end{eqnarray*}
        After establishing the key, the parties exchange the encrypted messages to verify the keys.

\section{Finite-blocklength sizing of the reliability code}\label{sec:Finite-blocklength sizing of the reliability code}
We model the channel between Alice and Bob as \(\texttt{BSC}_p(m)\), with bit error rate $p = 0.1666$. A channel with such an error probability has:
\begin{equation}
    C \approx 0.3501, \qquad
V \approx 0.7490. 
\end{equation}
Based on the normal approximation of Polyanskiy–Poor–Verdú, using the equation \eqref{eq:payload} we can estimate the reliably decodable payload size for considered channel. We assume blocklength \(n = 1990\) and block error probability as \(\epsilon = 10^{-5}\), then by \eqref{eq:payload}:
\begin{equation}
    \log M^*(n, \epsilon) \approx 537.59 \text{ bits}.
\end{equation}
Therefore, in an ideal case, we obtain the error correction code which encodes \(538\) bit into \(1990\) bit codeword, with probability of decoding with error \(\epsilon = 10^{-5}\). Decreasing the probability of error to \(\epsilon = 10^{-6}\), decreases payload to \(519\) bits.

The parties do not exchange syndromes. Only information relating to permutations is broadcasted, as described in Section \ref{sec:Quantum key exchange protocol}.

\subsection{Privacy amplification and final key length}
    We model eavesdropper’s Eve observation as \(\texttt{BSC}_p(m)\) with higher error rate than the Bob's. We assume that the probability of the eavesdropped bit mismatching Alice's bit is the same as that of mismatching Bob's bit:
    \begin{equation}
        p_{AE} = p_{BE} = p_{E} = 0.3333
    \end{equation}
    We estimate the secrecy capacity gap \cite{Yang2016}:
    \begin{equation}
        C_S = h(p_E) - h(p) \approx 0.9182958 - 0.6498676 \approx 0.2684282\ \text{bits/use}.
    \end{equation}
    According to \cite{Yang2016} we can estimate bound for the extractable secret key length:
    \begin{equation}
        \ell(k,\epsilon,\delta) \approx k C_s - \sqrt{kV} Q^{-1}(\epsilon) - \sqrt{kV_E} Q^{-1}(\delta) + \frac{1}{2} \log k,
    \end{equation}
    where \(V_E\) is dispersion for Eve's channel and \(\delta\) is information leakage.

    Using the same code parameters as for the channel between Alice and Bob, let us determine the key length. We assume \(\delta = 10^{-6}\), for the selected probability of error for Eve, her dispersion is \(V_E = 0.2222\). Then
    \begin{equation}
        \label{eq:key secrecy bound}
        \ell(1990,10^{-5},10^{-6}) \approx 275.04\ \text{bits}.
    \end{equation}

After establishing matching bits, in the final step of the protocol we instantiate privacy amplification using a 2-universal extractor based on Toeplitz hashing~\cite{Bennett_1995_GeneralizedPA,Impagliazzo_1989_LHL,Hayashi_2013_Toeplitz}. This extractor maps the decoded $n = 538$ bits to a final key of length 256 bits, leaving approximately $19$ bits of margin relative to the bound~\eqref{eq:key secrecy bound}.

The Toeplitz matrix has dimension $\ell \times n$ and can be specified by a public seed of length $\ell + n - 1$ bits~\cite{Smith_2023_Toeplitz,Abidin_2010_UniversalHash}.
The seed is chosen uniformly at random and may be sent in the clear, as it is independent of the payload and does not affect secrecy.

\subsection{Transmission length and theoretical efficiency}\label{subsec:Transmission length and theoretical efficiency}
As we described in Section \ref{sec:Protocol Description}, during transmission parties exchange random bits. Each party has random sequence of 1990 bits, Alice has \(\texttt{seed}_A^{ECC}\) and Bob has \(\texttt{seed}_B^{ECC}\). At each round, the parties randomly select whether the sampled bit comes from Alice's or Bob's sequence. The transmission ends when each party has sent at least as many 0s and 1s as there are in its sequence. 

So now we have to compute at which moment the protocol will stop in the average case. By the average case, we mean the average number of randomly sampled bits required to obtain proper sequences of minimal length. Below we present an empirical experiment used to determine the average transmission length.

We performed simulations using SageMath \cite{SageMath}, with the goal of:
\begin{itemize}
    \item presenting a general idea of the protocol with the quantum part simulated as a binary symmetric channel;
    \item determining the experimental overhead.
\end{itemize}

In order to determine the number of rounds required to establish our key, we performed a simulation without decoding. Decoding does not affect the length of the transmission. It takes place after each party has collected a sufficient number of proper bits. 

Parties exchange random sequences of length 1990 bits, as described in Section \ref{sec:Finite-blocklength sizing of the reliability code}. These sequences correspond to private components of keys encoded with ideal error correction code. The results are presented in Table \ref{tab:overhead}. With an ideal error correction code, parties exchange only one codeword. If that is the case, the number of overhead rounds $\Delta$ is equal to the number of overhead bits in communication, \(\delta = \Delta\). Parties needs to exchange two \(1990\)-bit sequences; therefore, the theoretical minimal number of rounds is \(3980\) if the parties always send proper bits. 
To measure the average number of rounds, we performed 10,000 trials and obtained an average equal to \(4149\). That means that, on average, the parties need to send only \(169\) bits more the in the best-case scenario. For the trials conducted, the smallest overhead was \(7\) and the highest number of rounds was \(4567\). 

We now move on to the estimation of the total overhead of bits needed to establish two 256-bit keys. Each key is encoded with an ideal correction code; the length of the codeword is 1990. In the ideal case, the protocol requires 3980 steps, incurring \(7.77\) qubits per bit. 

On the other hand, in the worst case scenario, the protocol will never finish. This will happen if the parties have a malicious random generator drawing a constant number. We disregard this case as not realistic.

As the experiment shows, the protocol always finished in at most 4567 rounds, requiring \(8.92\) qubits per bit. On average, the key managed to be exchanged after 4149 rounds, translating to \(8.1\) qubits per bit.
\begin{table}[h]
    \centering
    \begin{tabular}{|C{0.15\textwidth}|C{0.175\textwidth}|C{0.2\textwidth}|C{0.2\textwidth}|C{0.15\textwidth}|} \hline
         Number of trials & Minimal number of rounds & Maximal number of rounds & Average number of rounds & Standard deviation \\ \hline
    10000 & 3987 & 4567 & 4149 & 83.59 \\ \hline
    \end{tabular}
    \caption{Summary of the experiment estimating the required number of rounds}
    \label{tab:overhead}
\end{table}

\section{Sample implementation of the proposed QKD scheme}
In this section, we demonstrate how the processing and key agreement of our QKD scheme can be implemented in practice. 
While the previous section described the theoretical efficiency of the protocol -- yielding an overhead close to~8 -- here, we present a practical, working realization of the scheme.

The key difference is that we compose two error-correcting codes (ECCs) to compensate for the relatively high Bob's quantum-channel bit-error rate (BER) of~14.65\% (we can estimate the Bob's BER), see Section~\ref{subsec:Causal game setting}. In such scenario, the BER of Eve is around 50\%.
To do so efficiently, we use:
\begin{enumerate}
  \item a \(2\)-out-of-\(3\) majority-voting code (MVC), required to lower the raw BER. Errors will propagate only if two or three errors occur in a 3-bit word:
  \begin{equation}
      \text{BER'} = 3p^2(1-p) + p^3.
  \end{equation}
  The application of MVC reduces BER from 14.65\% to an effective 5.82\%;
  \item a BCH\((31,11)\) code\cite{channelcodes}, which encodes 11 input bits into 31 code-word bits and can correct up to five errors in each 31-bit block.
\end{enumerate}

One might ask why the BCH\((31,11)\) code alone is insufficient, since it corrects up to five errors and therefore works for BERs below \(5/31 \approx 16.13\,\%\). 
With a per-bit error probability \(p = 0.1465\) the probability that a 31-bit block contains at most five errors is
\[
\Pr(X\le5)
  =\sum_{k=0}^{5}\binom{31}{k}
    p^{k}(1-p)^{31-k}
  \approx 0.7027.
\]
Thus, a single block is decodable with probability only about \(70.27\,\%\). To transmit a 256-bit raw key we must encode at least 24 blocks (because \(24\times11=264 > 256\)). The probability that every block decodes correctly is then $0.7027^{24} \approx 0.021\%$, which is clearly negligible.

Therefore, our practical QKD prototype proceeds as follows:
\begin{enumerate}
  \item Generate a random 264-bit sequence \(r\).
  \item Divide \(r\) into 24 blocks of 11 bits and encode each block with the BCH\((31,11)\) code, producing a 744-bit sequence \(s\).
  \item Encode \(s\) with the MVC\((2,3)\), yielding a 2232-bit sequence \(s_1\).  
        In other words, the concatenated code is \(ECC_{2232,264}\).  
        From this point, the protocol follows the steps described in Section~\ref{sec:Quantum key exchange protocol}.
\end{enumerate}

Applying MVC\((2,3)\) reduces the effective BER to 5.82\%.  
The probability that a BCH block now contains at most five errors is
\begin{equation}
    \Pr(X\le5)
  =\sum_{k=0}^{5}\binom{31}{k}
    (0.0582)^{k}(0.9418)^{31-k}
  \approx 0.9919.
\end{equation}
Consequently, the probability that all 24 blocks decode correctly is $0.9919^{24} \approx 0.8227$.
Therefore, in theory, the single key part will be established with probability about \(82.27\%\). 
Finally, the probability that the entire protocol will be successful equals \(0.8227^2\approx 0.6768\).

To illustrate the practicality of our protocol, we conducted the following experiment. The simulation was performed using SageMath\cite{SageMath}:
\begin{itemize}
    \item we simulate Quantum Bit Exchange as binary symmetric channel as described in Section \ref{sec:modeling as BSC};
    \item we use BCH codes from inbuilt SageMath library;
\end{itemize}
The results of the experiment are presented in the Table \ref{tab:overhead 2}. As described earlier Each party wants to send a 264-bit fragment of the key. To do this, each party must send 2232 bits. Experiments show that establishing both session keys (528 bits in total) requires, on average, 5301 rounds, corresponding to an overhead of roughly~10 times. Since not every QKD session succeeds (about 86.7\% of sessions produce both session keys), the effective overhead is approximately~12.
\begin{table}[h]
  \centering
  \begin{tabular}{|C{0.25\textwidth}|C{0.15\textwidth}|C{0.15\textwidth}|C{0.15\textwidth}|C{0.15\textwidth}|}
    \hline
    Number of successful key exchanges &
    Minimum rounds &
    Maximum rounds &
    Average rounds &
    Standard deviation \\
    \hline
    867 & 5018 & 5642 & 5301 & 110.41 \\
    \hline
  \end{tabular}
  \caption{Experimental overhead for 867 successful key exchanges.}
  \label{tab:overhead 2}
\end{table}
It is worth noting that practical experiments bring a much higher probability of the success (86.7\%) than the theoretical estimations. The main reason for such behavior is a fact, that sometimes the codewords with more than 5 errors also may be decoded correctly. 

\section{Conclusions and outlook}\label{sec:Conclusions and outlook}
We have introduced a bipartite QKD protocol leveraging causally non-separable (CNS) processes. By harnessing the indefinite causal order intrinsic to such a resource, Alice and Bob can reliably share bits with about 85.35\% agreement per round in the absence of eavesdropping. Although this yields a relatively high raw BER (near 14.65\%), classical information theory reveals that error-correction codes can still operate efficiently within this range.

Future directions include:
\begin{itemize}
\item \emph{Experimental feasibility:} Realizing causally non-separable processes might involve sophisticated quantum networks or ancillary degrees of freedom enforcing indefinite causal order. One possibility for experimental realization could be achieved by using quantum switch~\cite{goswami2018indefinite} along with quantum processing. Quantum switch consists an example of casually non-separable process that has been experimentally implemented~\cite{stromberg2023demonstration} and for which the certification of working correctness can be achieved~\cite{van2023device}.

\item \emph{Device imperfections:} Incorporating technical problems as information loss, detector inefficiencies or side-channel attacks~\cite{Renner_2005} is an important step toward practical deployment.

\item \emph{Hybridization with post-quantum cryptography:} One can combine CNS-based QKD with classical post-quantum schemes, using a short quantum key to bootstrap or authenticate a larger post-quantum cryptography-based system, thereby maximizing overall security and flexibility.
\item \emph{Broader indefinite causal structure protocols:} Investigating whether other indefinite causal structures might produce higher correlations or reduce error rates.
\end{itemize}
By demonstrating how a CNS resource can be mapped to a QKD-like advantage, we hope to inspire further studies on the cryptographic potential of indefinite causal structures, complementing well-known entanglement-based QKD protocols.


\funding{PL is supported by the Ministry of Education, Youth and Sports of the Czech Republic through the e-INFRA CZ (ID:90254), with the financial support of the European Union under the REFRESH – Research Excellence For REgionSustainability and High-tech Industries project number CZ.10.03.01/00/22\_003/0000048 via the Operational Programme Just Transition. \\
RK is supported by the European Union under the Quantum error correction codes enhanced by reinforcement learning dedicated for the Ising model-based optimization, contract nr. 01906/2025/RRC via the Operational Programme Just Transition and Moravian-Silesian Region.
GRM acknowledges funding from the European Innovation Council accelerator grant COMFTQUA, no. 190183782.}

\roles{
RK (20\%) and PL (20\%) - QKD scheme development, quantum security analysis, article writing \newline
GRM (10\%) - article writing, literature review, performance comparison \newline
ML (25\%) and MW (25\%) - error correction strategy development, security analysis, article writing, project administration)
}


\data{The code that supports the findings of study is openly available on Github~\cite{solution}.}

\suppdata{}

\bibliographystyle{iopart-num}
\bibliography{bibliography}

\end{document}